\documentclass[letterpaper]{article} 
\usepackage{aaai2026}  
\usepackage{times}  
\usepackage{helvet}  
\usepackage{courier}  
\usepackage[hyphens]{url}  
\usepackage{graphicx} 
\usepackage{natbib}  
\usepackage{caption} 
\usepackage{algorithm}
\usepackage{algorithmic}
\usepackage{multicol, multirow, pifont, graphicx}
\usepackage{xcolor}
\usepackage{arydshln}
\usepackage{amsmath, amsfonts}
\usepackage{booktabs}

\usepackage{newfloat}
\usepackage{listings}
\DeclareCaptionStyle{ruled}{labelfont=normalfont,labelsep=colon,strut=off} 
\floatstyle{ruled}
\newfloat{listing}{tb}{lst}{}
\floatname{listing}{Listing}
\title{AdaptDiff: Adaptive Guidance in Diffusion Models for Diverse and Identity-Consistent Face Synthesis (Student Abstract)}
\author {
    Eduarda Caldeira\textsuperscript{\textrm 1},
    Tahar Chettaoui\textsuperscript{\textrm 1},
    Naser Damer\textsuperscript{\textrm 1, \textrm 2},
    Fadi Boutros\textsuperscript{\textrm 1}
}
\affiliations {
    \textsuperscript{\rm 1}Fraunhofer IGD\\
    \textsuperscript{\rm 2}Department of Computer Science, TU Darmstadt
}

\usepackage{bibentry}

\begin{document}

\maketitle

\begin{abstract}
\vspace{-3mm}
Diffusion models conditioned on identity embeddings enable the generation of synthetic face images that consistently preserve identity across multiple samples. Recent work has shown that introducing an additional negative condition through classifier-free guidance during sampling provides a mechanism to suppress undesired attributes, thus improving inter-class separability. Building on this insight, we propose a dynamic weighting scheme for the negative condition that adapts throughout the sampling trajectory. This strategy leverages the complementary strengths of positive and negative conditions at different stages of generation, leading to more diverse yet identity-consistent synthetic data. 
\end{abstract}

\begin{links}
    \link{Code}{https://github.com/EduardaCaldeira/NegFaceDiff/}
\end{links}
\section{Introduction}
The use of synthetic data generated by diffusion models (DMs) to train face recognition (FR) has significantly evolved in recent years due to the need to address privacy and data protection issues associated with the use of privacy-sensitive authentic datasets. 
To enable FR training with state-of-the-art (SOTA) losses \cite{wang2018cosfacelargemargincosine}, i.e., margin-penalty softmax losses, a training dataset with class (identity) labels is required. Recent works leverage identity-conditional DMs to generate such labeled datasets.
Inspired by the success of negative prompting in text-to-image generation, recent work has shown that including an extra negative condition to guide the diffusion process towards the exclusion of undesired characteristics improves FR performance while increasing intra-class variability and inter-identity variation \cite{caldeira2025negfacediff}. In this work, we prove that the benefits of sampling with negative conditions are hindered by the fixed nature of their magnitudes, as strong negative conditions restrain the DM's freedom to explore the sampling space. We propose to adapt the negative condition strength during the sampling process, enabling higher freedom in the early stages, when little to no identity information is encoded and the latent space should be freely explored, and taking advantage of the benefits of negative prompting at later sampling stages, which can operate in a more restricted space.

\section{Methodology}
DMs are trained to learn how to approximate an input (reverse process) by decoding noise extracted from it (forward process). A trained DM can generate a synthetic image by denoising a random noise seed. To generate several facial images of the same identity (class label), DMs use an identity condition $p^+$. A set of samples belonging to the same identity can be sampled by fixing $p^+$ and varying the initial random noise seed \cite{DBLP:conf/iccv/BoutrosGKD23}. 
An extra negative identity condition $p^-$ can also be used during sampling \cite{caldeira2025negfacediff}, resulting in additional guidance toward the exclusion of undesired characteristics in generated data, and higher separation between the identities, improving FR performance. This is achieved by incorporating the effects of the predicted noise added at time step $t$, $\epsilon_\theta(\mathbf{x}_t,t,p), \; p \in\{p^+, p^-\}$, for both $p^+$ and $p^-$: $\hat{\epsilon}_\theta(x_t, t, p^+, p^-) = (1 + w) \, \epsilon_\theta(x_t,t,p^+) - w \, \epsilon_\theta(x_t,t,p^-),$
where $w$ is the guidance strength of classifier's free guidance (CFG) negative condition \cite{ban2024understanding, DBLP:conf/icml/Wang0HG24}. The subtraction ensures that generated images follow $p^+$ and deviate from $p^-$. 

\begin{table}
\centering
\resizebox{\linewidth}{!}{
\begin{tabular}{@{}cc|cccccc@{}}
\textbf{Method} & \textbf{$w$ \& $w_{max}$} & \textbf{LFW} & \textbf{AgeDB}  & \textbf{CAFLW} & \textbf{CFPFP}  & \textbf{CPLFW}  & \textbf{Avg}\\
\hline
IDiff-Face  & 0 &  96.30	 & 78.15    &  86.23 & 81.51 & 77.57     & 83.95 \\
\hline
\multirow{2}{*}{NegFaceDiff} & 0.5 & \textbf{96.78} & \textbf{81.35} & \textbf{87.67} & \textbf{81.89} & \textbf{78.43} & \textbf{85.22} \\
{} & 1.0 & 92.82 & 73.98 & 80.22 & 74.81 & 72.10 & 78.79 \\
\hline 
\multirow{2}{*}{AdaptDiff} & 0.5 & 96.88 & 79.33 & \textbf{82.70} & 87.40 & 78.53 & 84.97 \\
{} & 1.0 & \textbf{97.00} & \textbf{81.57} & 82.29 & \textbf{88.17} & \textbf{79.08}	& \textbf{85.62} \\
\hline
\end{tabular}}
\caption{Accuracies (in $\%$) of FR models trained on NegFaceDiff and our AdaptDiff with different $w$ \& $w_{max}$.}
\label{tab:w_ablation}
\end{table}

\begin{figure*}
    \centering
    \includegraphics[width=0.9\linewidth]{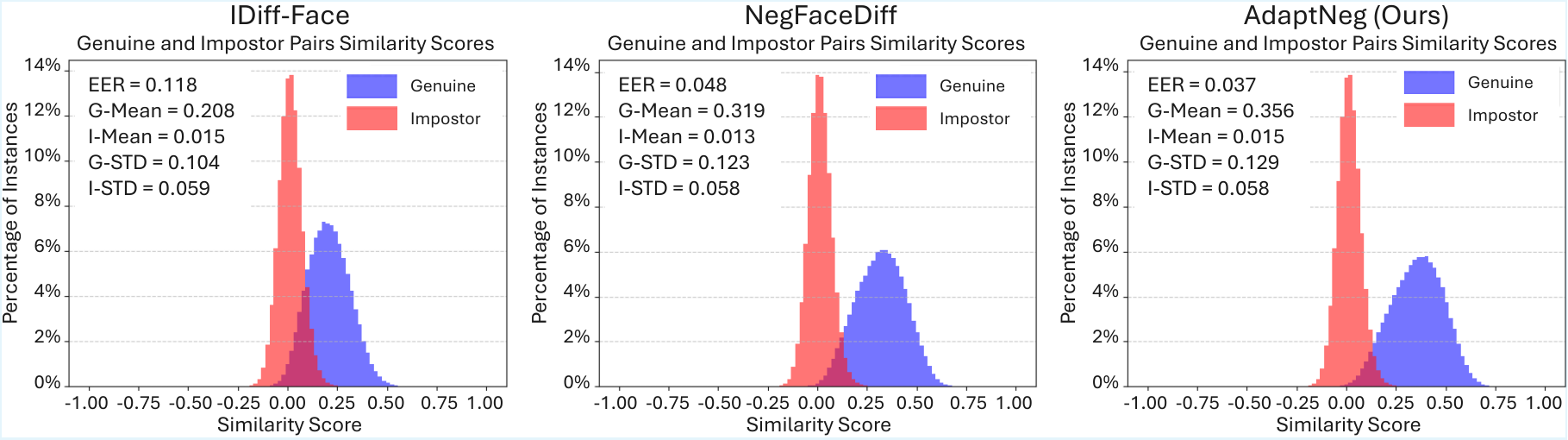}
    \caption{Histograms of genuine and impostor score distributions of SOTA baselines and AdaptDiff trained on FFHQ.
    }
    \label{fig:dist}
\end{figure*}

\begin{figure}
    \centering
    \includegraphics[width=0.9\linewidth]{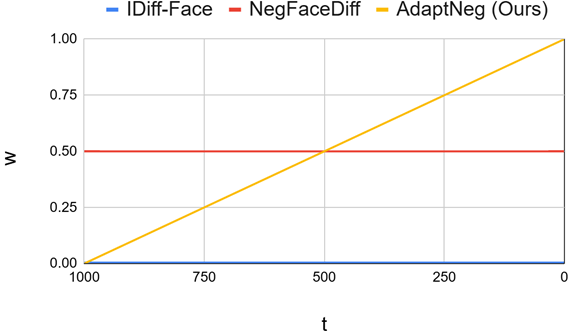}
    \caption{$w$ as a function of the sampling time step, $t$. \textbf{Blue:} IDiff-Face fixes $w=0$. \textbf{Red:} NegFaceDiff fixes $w=0.5$. \textbf{Yellow:} AdaptDiff adapts $w$ during sampling ($w=1-t/T$).}
    \label{fig:scheduler}
\end{figure}

\begin{table}[ht!]
\centering
\resizebox{\linewidth}{!}{%
\begin{tabular}{c|cccccc|cc}
    \multirow{2}{*}{\textbf{Method}} &
  \multirow{2}{*}{\textbf{LFW}} &
  \multirow{2}{*}{\textbf{AgeDB}} &
  \multirow{2}{*}{\textbf{CFPFP}} &
  \multirow{2}{*}{\textbf{CALFW}} &
  \multirow{2}{*}{\textbf{CPLFW}} &
  \multirow{2}{*}{\textbf{Avg}} &
  \multicolumn{2}{c}{\textbf{IJB-C}} \\
  &&&&&&&
  \textbf{$10^{-5}$} &
  \textbf{$10^{-4}$} \\ \hline
 
 C-WF &
  99.55 &
  94.55 &
  95.31 &
  93.78 &
  89.95 &
  94.63 & 93.96 & 96.05
  \\ \hline
  Arc2Face & 98.81 & 90.18 & 91.87 & 92.63 & 85.16 & 91.73 & - & - \\  \hline \hline
 IDiff-Face*    & 96.48 & 77.27 &	81.71 &	86.95 &	78.55 &	84.19 & 16.85 &	37.58\\  
           NegFaceDiff*    & 96.80 & 80.45	& 82.34 & 87.32 &	78.62 &	85.11 & \textbf{23.88} & 49.13 \\ 
       \textbf{AdaptDiff*}      & \textbf{96.97} &	\textbf{81.08} &	\textbf{83.39} &	\textbf{87.58} &	\textbf{79.77} &	\textbf{85.76} &		23.77 &	\textbf{55.81}\\ \cdashline{1-9}
            ID$^3$                & 97.28 & 83.78 & 85.00 & 89.30 & 77.13 & 86.50 & - & - \\
 IDiff-Face   & \textbf{98.00} & 86.43 & \textbf{85.47} & \textbf{90.65} & 80.45 & 88.20 & 20.60 & 62.60 \\  
         NegFaceDiff  & 97.60 &	86.53 &	85.33 &	90.28 &	80.73 &	88.10 & 58.09 & 73.93 \\ 
      \textbf{AdaptDiff}     & 97.62 & \textbf{86.63} & 85.04 & 90.43 & \textbf{81.82} & \textbf{88.31} & \textbf{62.38} & \textbf{76.56} \\ \hline \hline
    IDiff-Face*  & \textbf{98.92} &	89.40 &	90.54 &	90.92 &
  86.60 &	91.28 &	73.18 & 82.71\\ 
         NegFaceDiff*  & \textbf{98.92} & 89.23 & 90.59 & 91.13 & 87.88 &	91.55 & \textbf{76.81} & 85.13 \\
         \textbf{AdaptDiff*} & 98.82 & \textbf{89.77} & \textbf{91.14} & \textbf{91.62} & \textbf{88.15} & \textbf{91.90} & 76.68 & \textbf{85.48}\\  \cdashline{1-9}
   ID$^3$            & 97.68 & \textbf{91.00} & 86.84 & 90.73 & 82.77 & 89.80 & - & - \\
   DCFace    & 98.55 & 89.70 & 85.33 & 91.60 & 82.62 & 89.56 & 60.80 & 74.63 \\ 
    IDiff-Face    & 98.98 & 89.30 & 90.97 & 91.25 & 86.87 & 91.47 & 23.44 & 69.69\\ 
         NegFaceDiff      & \textbf{98.98} &	90.02 &	91.67 &	\textbf{91.65}	& \textbf{88.82} &	\textbf{92.23} & \textbf{77.38}	& 86.11 \\
          \textbf{AdaptDiff}  & 98.83 &	89.75 & \textbf{91.77} & 91.42 &	88.77 & 92.11 &		76.85 & \textbf{86.48} \\ 
\hline
\end{tabular}%
}
\caption{Accuracies (in \%) of FR models trained on AdaptDiff and SOTA. ``*'': FR results obtained without data augmentation. ``-'': works that did not release pre-trained models/generated data and did not report these results. Arc2Face was trained on WebFace4M; the models in the second and third blocks were trained on FFHQ and C-WF, respectively.}
\label{tab:sota}
\end{table}

In this work, we 
dynamically adapt $w$ during sampling, instead of fixing its value, due to 
the varying role of identity information across time steps. At early sampling ($t \xrightarrow{} T$), 
$x_t$ contains little to no identity information. Thus, the DM should explore the identity space with a high degree of freedom and weighting $p^-$ with strong magnitude 
will unnecessarily hinder intra-class variations \cite{DBLP:conf/iclr/SadatBBHW24}.
At later steps ($t \xrightarrow{} 0$), $x_t$ incorporates relevant identity features, and weighting $p^-$ with a strong magnitude further pushes the generated samples towards $p^+$, ensuring high inter-class separability. Hence, progressively increasing $w$ with a linear schedule ($w = w_{max} \times (1 - \frac{t}{T})$, as shown in Figure \ref{fig:scheduler}) facilitates earlier steps while still taking advantage of the improvements brought by negative conditions. 
\section{Results}
We conduct studies on two IDiff-Face versions, pre-trained on either Flickr-Faces-HQ (FFHQ) or CASIA-WebFace (C-WF), following \cite{caldeira2025negfacediff}.

\noindent \paragraph{$w$ \& $w_{max}$ Ablation Study:}
NegFaceDiff's performance significantly drops from $w=0.5$ to $w=1.0$ (Table \ref{tab:w_ablation}). 
Since $p^-$ pushes sampling towards the desired identity, highly accentuating it through the whole sampling unnecessarily hinders intra-class variability, restricting fixed negative sampling to a lower $w$. AdaptDiff addresses this limitation by reducing the guidance strength magnitude in early sampling, allowing to push the negative condition strength to higher magnitudes in later stages without compromising performance. This is supported by AdaptDiff's superiority with $w_{max}=1.0$ over $w_{max}=0.5$. 
Hence, we set $w_{max}=1.0$ for all the remaining experiments.

\noindent \paragraph{Identity Separability:}
We compare the identity separability of the data generated by IDiff-Face, NegFaceDiff and our AdaptDiff, following \cite{caldeira2025negfacediff}. Figure \ref{fig:dist} displays their genuine and impostor score distributions and relevant metrics. AdaptDiff presents the smallest overlap between the two distributions, revealing higher identity-separability, as numerically supported by its lower EER. Moreover, AdaptDiff has the highest genuine standard deviation, suggesting higher intra-identity variation.

\noindent \paragraph{Face Recognition:}
Table \ref{tab:sota} presents evaluation results of FRs trained on AdaptDiff and previous DM SOTA. When using a DM pre-trained on FFHQ, AdaptDiff surpassed SOTA approaches on average on the small-scale benchmarks and significantly outperformed them on the challenging IJB-C in most scenarios. Using data augmentation in FR training significantly improved FR performance. When using a DM trained on C-WF, AdaptDiff presented very competitive performance with SOTA. Data augmentation slightly improved FR performance. The lack of significant improvements, such as those verified with FFHQ, derives from the higher intra-identity variation of data generated by DMs trained on C-WF, which reduces the impact of strategies tailored to improve this property, such as AdaptDiff.

\section{Conclusion}
This work proposed a novel adaptive strategy to weight CFG positive and negative conditions across sampling time steps for identity-conditioned DMs. AdaptDiff provides a trade-off between the DM's freedom to explore its latent space and the benefits introduced by negative prompting, resulting in synthetic datasets with higher inter-class separability and intra-class variation, and improving FR performance.


\section{Ethics Statement}

Face recognition (FR) systems are currently deployed in several applications, ranging from automated border control \cite{DBLP:journals/cviu/GuoZ19,abs_gate} to biometric on-device authentication \cite{prakash2021biometric}. While these models were traditionally trained on authentic facial data, this practice has raised several concerns regarding user consent, privacy and general data protection, along with the technical limitations of collecting, sharing and storing sensitive biometric data as well as the respective legal and ethical concerns \cite{GDPR_practice, bipa}. To mitigate this issue, recent works \cite{arc2face, DBLP:conf/nips/Xu0WXDJHM0DH24, SyntheticFRSurvay, DBLP:conf/cvpr/Kim00023} focused on enhancing facial data generation models, towards the generation of synthetic FR training datasets. Our AdaptDiff constitutes one of such research efforts, bringing improvement over previous SOTA synthetic face generation approaches. We further acknowledge but strongly reject any potential for unlawful or ill-intended use of the developed sampling technology. Although we exclusively use DMs to generate faces from synthetic identities in the work, DMs can also generate synthetic pictures of real identities \cite{arc2face}, which can be used for malicious purposes such as impersonation and evidence fabrication.

\section{Acknowledgments}

This research work has been funded by the German Federal Ministry of Education and Research and the Hessian Ministry of Higher Education, Research, Science and the Arts within their joint support of the National Research Center for Applied Cybersecurity ATHENE.


\bibliography{aaai2026} 

\newpage

\section{Supplementary Material}
This supplementary material presents additional results that complement those in the main document, as well as detailed descriptions of the experimental setups and evaluation benchmarks.
This includes: 
\begin{itemize}
    \item A detailed description of the experimental setup
    \item Complementary identity separability evaluation results
\end{itemize}

\section{Experimental Setup}

This section presents detailed information of the experimental setup followed in this work. We start by describing the pre-trained DMs used to generate the AdaptDiff datasets. We then elaborate on the two sampling methods used in different experimental phases. We further present detailed descriptions of the conducted identity separability evaluation and AdaptDiff's FR training setup. Finally, we describe the evaluation benchmarks and datasets used to assess FR performance.


\paragraph{Pre-Trained Diffusion Models:} 

The pre-trained DM used as a base for this study was IDiff-Face \cite{DBLP:conf/iccv/BoutrosGKD23} with a contextual partial dropout of 25\%. We conduct studies on two versions of this model trained on different datasets, following \cite{caldeira2025negfacediff}: FFHQ and C-WF. Our choice for IDiff-Face \cite{DBLP:conf/iccv/BoutrosGKD23} is due to its simplicity, to its wide adaption in many solutions submitted to several international competitions \cite{DBLP:journals/inffus/MelziTVKRLDMFOZZYZWLTKZDBVGFFMUG24,DBLP:conf/fgr/Otroshi-Shahreza24,DEANDRESTAME2025103099}, to its SOTA performances (e.g., winner solution in SDFR competition \cite{DBLP:conf/fgr/Otroshi-Shahreza24}) and its previous adaption to include negative conditions in sampling \cite{caldeira2025negfacediff}, providing an extra baseline for comparison. 

\paragraph{Sampling Methods:}

We extended the best-performing sampling mechanism described by NegFaceDiff \cite{caldeira2025negfacediff}, Far-Neg, by adapting the value of the weighting condition $w$ during the sampling process. Note that both our baselines \cite{DBLP:conf/iccv/BoutrosGKD23, caldeira2025negfacediff} utilized DDPM \cite{Ho2020} with 1000 steps for training and sampling. In the ablation study described in the main document, we opt to use a Denoising Diffusion Implicit Model (DDIM) \cite{DBLP:conf/iclr/SongME21} with 200 steps to perform the sampling phase. Since DDIM improves the sampling speed of DDPM while maintaining the sample quality to a large degree, it allows to speed up the ablation process while ensuring that its results do not lose significance. The SOTA results provided in the main paper are reported using DDPM to generate 10k novel identities with 50 images per identity, to provide a fair comparison with previous works.


\begin{figure*}
    \centering
    \includegraphics[width=\linewidth]{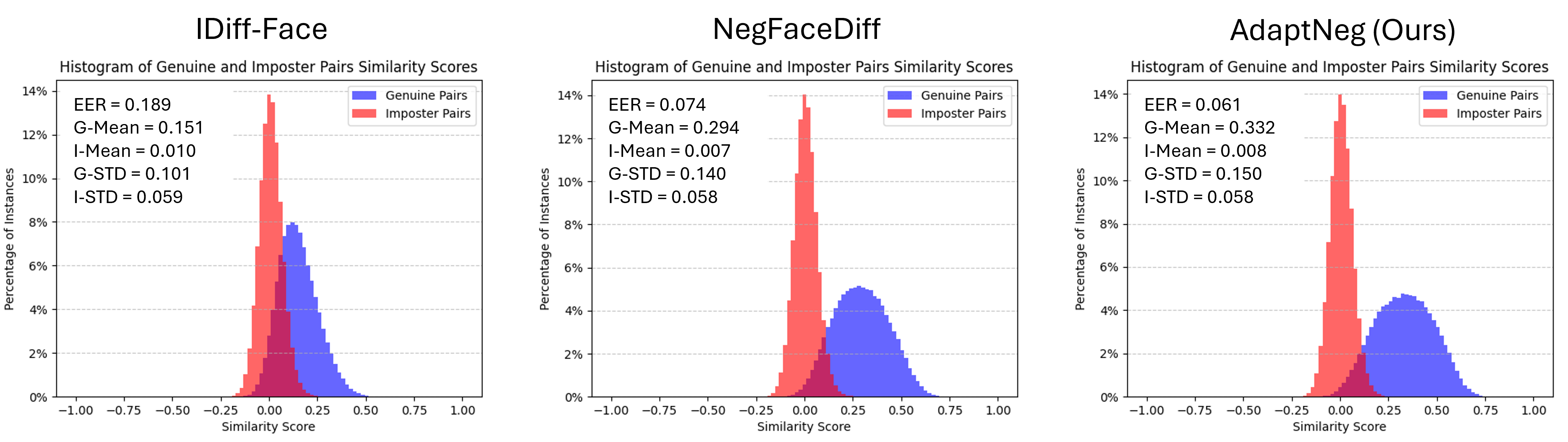}
    \vspace{-5mm}
    \caption{Histograms of genuine and impostor score distributions of SOTA baselines and AdaptDiff trained on C-WF.}
    \label{fig:dist_CASIA}
\end{figure*}

\tabcolsep=0.1cm
\begin{table*}[t!]
\centering
\caption{Evaluation of the identity-separability of datasets generated by IDiff-Face, NegFaceDiff and our AdaptDiff based on DMs trained on FFHQ (second block) and C-WF (third block). The first block present the results on authentic C-WF and LFW, which are provided as a reference. All the synthetic evaluations are based on synthetically generated datasets with $1{,}000$ identities and $20$ sampled images per identity. The lowest errors and the highest genuine-imposter separability scores (FDR) on synthetic datasets are marked in \textbf{bold}.}
\vspace{-2mm}
\resizebox{\linewidth}{!}{%
    \begin{tabular}{@{}cccccccccccccc@{}}\toprule
    \multicolumn{1}{c}{\textbf{ }} & \phantom{abc} &  \multicolumn{1}{c}{\textbf{ }} & \phantom{abc} & \multicolumn{3}{c}{\textbf{Operation Metrics}} & \phantom{abc} & \multicolumn{5}{c}{\textbf{Score Distributions}}\\
    \cmidrule{9-14}
    \multicolumn{1}{c}{\textbf{ }} & \phantom{abc}  & \multicolumn{5}{c}{\textbf{ }}& \phantom{abc} & \multicolumn{2}{c}{genuine} &   \multicolumn{2}{c}{imposter} &  \multicolumn{1}{c}{\textbf{ }}\\
    
    \cmidrule{1-1} \cmidrule{3-3} \cmidrule{5-7} \cmidrule{9-14}
    \multicolumn{1}{c}{\textbf{DM Training Dataset}} && \textbf{Method} && EER $\downarrow$  & FMR100 $\downarrow$  & FMR1000 $\downarrow$  && mean & std & mean & std & FDR $\uparrow$ \\
    \midrule
 - && C-WF \cite{DBLP:journals/corr/YiLLL14a} && 0.076  & 0.092  & 0.107   &&  0.536 &  0.215  &  0.003  &  0.070  &  5.541 \\
    - && LFW \cite{LFWDatabase} &&  0.002   &  0.002  &  0.002  &&  0.708  &  0.099  &  0.003  &  0.070  &  33.301 \\
    \cmidrule{1-1} \cmidrule{3-3} \cmidrule{5-7} \cmidrule{9-14}%
    \multirow{3}{*}{FFHQ} && IDiff-Face \cite{DBLP:conf/iccv/BoutrosGKD23} && 0.118 & 0.341 & 0.562 && 0.208 & 0.104 & 0.015 & 0.059 & 2.631 \\
    && NegFaceDiff \cite{caldeira2025negfacediff} && 0.048 & 0.105 & 0.229 && 0.319 & 0.123 & 0.013 & 0.058 & 5.052 \\
    && \textbf{AdaptDiff (Ours)} && \textbf{0.037} & \textbf{0.074} & \textbf{0.173} && 0.356 & 0.129 & 0.015 & 0.058 & \textbf{5.806} \\
    \cmidrule{1-1} \cmidrule{3-3} \cmidrule{5-7} \cmidrule{9-14}
    \multirow{3}{*}{C-WF} && IDiff-Face \cite{DBLP:conf/iccv/BoutrosGKD23} && 0.189 & 0.558 & 0.767 && 0.151 & 0.101 & 0.010 & 0.059 & 1.454\\
    && NegFaceDiff \cite{caldeira2025negfacediff} && 0.074 & 0.170 & 0.314 && 0.294 & 0.140 & 0.007 & 0.058 & 3.579 \\
    && \textbf{AdaptDiff (Ours)} && \textbf{0.061} & \textbf{0.126} & \textbf{0.238} && 0.332 & 0.150 & 0.008 & 0.058 & \textbf{4.081} \\
    \bottomrule
    \end{tabular}
}
\label{tab:id_sep}
\end{table*}

\paragraph{Identity Separability Evaluation:}
\label{sec:setup_idsep}
The identity separability of the synthetic data generated by the considered baselines \cite{DBLP:conf/iccv/BoutrosGKD23, caldeira2025negfacediff} and our AdaptDiff is evaluated in the main paper for the datasets generated with a DM pre-trained on FFHQ and extended in this supplementary material for C-WF. The verification performances are reported as FMR100 and FMR1000, which are the lowest false non-match rate (FNMR) for a false match rate (FMR)$\leq$1.0\% and $\leq$0.1\%, respectively, along with the Equal Error Rate (EER). The histograms of genuine and imposter score distributions are also presented. We further report the genuine and imposter means (G-mean and I-mean) and standard deviation (G-STD and I-STD) to evaluate the datasets' inter-class separability and intra-class compactness, respectively. We also report the Fisher Discriminant Ratio (FDR) \cite{poh2004study} to quantify the separability of genuine and impostor scores. We used a pre-trained ResNet100 \cite{He2015DeepRL} with ElasticFace \cite{ElasticFace} on MS1MV2 \cite{Deng_2022,guo2016ms} to extract the feature representations needed for the evaluation described above. All the results provided are relative to datasets sampled with DDPM.

\paragraph{AdaptDiff FR Training Setup:}
\label{sec:fr_eval}

The FR models described in this work follow the setup described in \cite{caldeira2025negfacediff}. All the models used the ResNet50 \cite{He2015DeepRL} architecture and were trained with with the CosFace loss \cite{wang2018cosfacelargemargincosine}, with a magin penalty of 0.35 and a scale factor of 64. The mini-batch size is defined as 512 and the training process uses a Stochastic Gradient Descent optimizer with a momentum of 0.9 and a weight decay of 5e-4. The models are trained on our synthetically generated datasets and compared to the models trained with data sampled with the considered baselines. All these datasets contain 10k identities with 50 images per identity, totaling 500k samples. Random augmentation with 4 operations and a magnitude of 16 was applied during training, following \cite{ExFaceGAN}, unless otherwise explicitly noted.

\paragraph{Evaluation Benchmarks and Metrics:}
\label{sec:eval_benchmarks}
The FR models' performance of the models described in this work is evaluated and reported as the verification accuracy on five benchmarks, Labeled Faces in the Wild (LFW) \cite{LFWDatabase}, AgeDb-30 \cite{AgeDB30Database}, Cross-Age LFW (CA-LFW) \cite{CALFWDatabase}, Celebrities in Frontal-Profile in the Wild (CFP-FP) \cite{CFPFPDatabase}, and Cross-Pose LFW (CP-LFW) \cite{CPLFWDatabase}, following their official evaluation protocol. We further evaluated these models on the large-scale evaluation benchmark IARPA Janus Benchmark–C (IJB-C) \cite{DBLP:conf/icb/MazeADKMO0NACG18}. In this setting, we considered the official 1:1 mixed verification protocol and reported the verification performance as True Acceptance Rates (TAR) at False Acceptance Rates (FAR) of 1e-4 and 1e-5.

\section{Results}
This section presents results complementary to those analyzed in the main paper, namely a detailed identity separability analysis of datasets generated by the considered baselines and our AdaptDiff.

\paragraph{Identity Separability Evaluation:}
We extend the identity separability evaluation presented in the main paper to include the datasets generated by IDiff-Face, NegFaceDiff and our AdaptDiff when using a DM pre-trained on C-WF to perform sampling. The resultant genuine and impostor score distributions are presented in Figure \ref{fig:dist_CASIA}. AdaptDiff results in the smallest overlap between genuine and impostor distributions and the lowest EER, revealing higher identity-separability. Simultaneously, it achieves the highest genuine standard deviation, suggesting higher intra-identity variation. These results follow the same trends as those presented in the main paper for FFHQ-based DMs, showing that our AdaptDiff improves identity-separability and enhances intra-identity variation across sampling settings. Table \ref{tab:id_sep} presents a complete quantitative evaluation of relevant distribution metrics for both FFHQ and C-WF. When comparing our AdaptDiff in these two scenarios, using a DM trained on C-WF results in higher genuine standard deviation, suggesting higher intra-identity variation. A similar observation can be made for NegFaceDiff, while datasets sampled with IDiff-Face present comparable genuine standard deviation values. When sampling from a DM trained on FFHQ, the improvement in FR performances is more significant, in comparison to the case when we sampled from a DM trained on C-WF, since our strategy is specifically tailored to improve a property (intra-identity variation) that is already enhanced by default in this setting.  
In particular, the smaller performance improvement verified when using data augmentation to train the FR model in the AdaptDiff dataset sampled from a DM trained on C-WF can also be attributed to this property.

\end{document}